\documentstyle[11pt]{article}

\begin{document}
\begin{center}
{\Large \bf On the Possibility of Abnormally Intense Radiation Due to}

\vspace{3mm}

{\Large \bf the Rotation of Electron Around a Dielectric Sphere}

\vspace{7mm}

L. Sh. Grigoryan\footnote{E-mail: levonshg@iapp.sci.am}, H. F. Khachatryan,
S. R. Arzumanyan

\bigskip

{\em Institute of Applied Problems in Physics \\[0pt]
25 Nersessian St., 375014 Yerevan, Armenia}
\end{center}

\bigskip
\noindent {\bf Abstract}

\noindent
The abnormally intense radiation due to the uniform rotation of electron
around the equatorial plane of a dielectric sphere is obtained. It takes
place when the sphere surface is at a specific distance from the
electron orbit and when the Cherenkov condition for electron and the
matter of the sphere is satisfied.

\bigskip


\bigskip

\section{Introduction}

A number of important electromagnetic processes is conditioned by the
matter: the Vavilov-Cherenkov radiation, the X-ray transition
radiation, the radiation of channeled particles \cite{Bol} - \cite{Akhiez}.
In this connection it is of interest to study an influence of the matter on
the radiation of the relativistic charge rotating along a circle in a
permanent magnetic field (synchrotron radiation \cite{Sokol, Tern}).

The synchrotron radiation in an infinite uniform medium was studied in
\cite{Tsyt} and further in \cite{Zrel, Kit}. The radiation of a
nonrelativistic particle rotating uniformly around a dielectric sphere,
and the radiation of the particle rotating in close proximity to the ideally
conducting sphere were considered in the \cite{Magom}. In \cite{Arzum, ArzKot}
the expressions were obtained for the spectral and spectral-angular
distribution of the radiation intensity without restrictions on the
orbit radius and velocity of a particle rotating around a sphere with an
arbitrary dielectric permittivity.

In the present paper an analysis of the numerical calculations by the
formulae obtained in \cite{Arzum, ArzKot} is carried out. The peculiarities
of the radiation conditioned by the matter of a sphere and by its size, are
revealed.

\section {Basic formulae}

We present the basic formulae describing the radiation of a particle
with the charge $q$ and velocity $v=\omega_{e}r_{e}$ uniformly rotating
around a sphere in its equatorial plane ($r_{e}$ is the radius of
orbit). The magnetic permeability of the sphere we take equal to unity
and consider its dielectric permittivity $\varepsilon_{0}$ as an
arbitrary real quantity (we do not take into account the effects
connected with the radiation absorption), the sphere radius $r_{o}<r_{e}$.
The radiation intensity at the frequency $\omega=k\omega_{e}$ (after an
averaging over the rotation period $2\pi/\omega_{e}$) is determined by
the expression
         \begin {equation}
         \label{1}
           I_{k}=2\frac{q^{2}\omega_{e}^{2}}{c\sqrt{\varepsilon_{1}}}
           \sum_{s=0}^{\infty}(\mid{a_{kE}(s)}\mid^{2}+\mid{a_{kH}(s)}
           \mid^{2}),
        \end {equation}
where $\varepsilon_{1}$ is the dielectric permittivity of a medium
surrounded the sphere,
         \begin {eqnarray}
         \label{2}
           a_{kE}=kb_{l}(E)P_{l}^{k}(0)\sqrt{\frac{(l-k)!}{l(l+1)(2l+1)
           (l+k)!}}\,, \qquad l=k+2s, \nonumber\\
           a_{kH}=b_{l}(H)\sqrt{\frac{(2l+1)(l-k)!}{l(l+1)(l+k)!}}\cdot
           \frac{dP_{l}^{k}(y)}{dy}, \qquad y=0, \qquad l=k+2s+1
         \end {eqnarray}
are the dimensionless amplitudes describing the contributions of
multipole of the electric and magnetic kinds, respectively. In Eq.(\ref{2})
$P_{l}^{k}(y)$ are the associated Legendre polynomials, and $b_{l}$ is a
factor depending on
$k$, $x=r_{0}/r_{e}$, $\varepsilon_{0}$ and $\varepsilon_{1}$:
         \begin{eqnarray}
            b_{l}(H)=iu_{1}\left[j_{l}(u_{1})-h_{l}(u_{1})\frac{\lbrace j_{l}
            (\underline{x}u_{0}),j_{l}(\underline{x}u_{1})\rbrace }{j_{l}
            (xu_{0})h_{l}(xu_{1})}\right] ,
            \qquad u_{i}=k\sqrt{\varepsilon_{i}}\frac{v}{c}, \nonumber \\
            b_{l}(E)=(l+1)b_{l-1}(H)-lb_{l+1}(H)+\frac{1}{x^{2}}\left(\frac{1}
            {\varepsilon_{0}}-\frac{1}{\varepsilon_{1}}\right)\times
            \nonumber \\
            \times\left[j_{\underline{l-1}}(xu_{0})+j_{\underline{l+1}}
            (xu_{0})\right]\left[h_{\underline{l-1}}(u_{1})+h_{\underline{l+1}}
            (u_{1})\right]\frac{l(l+1)u_{0}j_{l}(xu_{0})}{lz_{l-1}^{l}+(l+1)
            z_{l+1}^{l}} ,
         \label{3}
         \end{eqnarray}
where
$h_{l}(y)=j_{l}(y)+in_{l}(y)$; $j_{l}$ and $n_{l}$
are the spherical Bessel and Neumann functions, respectively. In Eq.(\ref{3}) the following notations
are introduced:
        \begin{eqnarray}
        \label{4}
           \lbrace a(\underline{x}u_{i}),b(\underline{x}u_{j})\rbrace =a
           \cdot\frac{\partial{b}}{\partial{x}}-\frac{\partial{a}}{
           \partial{x}}\cdot b,\qquad f_{\underline{l}}(y)=\frac{f_{l}(y)}
           {\lbrace j_{l}(\underline{x}u_{0}),h_{l}(\underline{x}u_{1})
           \rbrace }\,,\nonumber\\ z_{\nu}^{l}=\frac{u_{1}j_{\nu}(xu_{0})
           h_{l}(xu_{1})/\varepsilon_{1}-u_{0}j_{l}(xu_{0})h_{\nu}(xu_{1})/
           \varepsilon_{0}}{u_{1}j_{\nu}(xu_{0})h_{l}(xu_{1})-u_{0}j_{l}
           (xu_{0})h_{\nu}(xu_{1})}.
        \end{eqnarray}
The derivation of Eq.(\ref{1}) is given in \cite{Arzum, ArzKot}.

In the case of homogeneous medium
($\varepsilon_{0}=\varepsilon_{1}=\varepsilon$)
        \begin {eqnarray}
        \label{5}
           b_{l}(H)=iuj_{l}(u),\qquad u=k\sqrt{\varepsilon}\frac{v}{c}\,,
           \nonumber\\b_{l}(E)=iu(2l+1)\left[j_{l}^{'}(u)+\frac{1}{u}
           j_{l}(u)\right],
        \end {eqnarray}
and therefore Eq.(\ref{1}), naturally, does not depend on $x$. One can also be
convinced that Eq.(\ref{1}) is transformed into the known formula \cite{Zrel,
Sokol, Tsyt, Kit, Land}
        \begin {equation}
        \label{6}
           I_{k}=kvq^{2}\frac{\omega_{e}^{2}}{c^{2}}\left[2J_{2k}^{'}
           (2k\beta\sqrt{\varepsilon})+(1-\frac{1}{\varepsilon\beta^{2}})
           \int\limits_{0}^{2k\beta\sqrt{\varepsilon}}J_{2k}(y)dy\right],
        \end {equation}
where $\beta=v/c$, $J_{k}(y)$ is the integer-order Bessel function,
and $\varphi^{'}(y)=d\varphi/dy$.

\section {Results of numerical calculations}

In Fig.1 along the axis of ordinates we plotted an average number of
electromagnetic field quanta
        \begin{equation}
        \label{7}
           n_{k}=\frac{2\pi I_{k}}{k\hbar\omega_{e}^{2}},
        \end{equation}
radiated per one period of rotation of electron with the energy 2 MeV
(the logarithmic scale), and along the axis of abscissa an order of
radiated harmonic in the range $1\leq{k}\leq50$ is plotted. The function
$n_{k}$ is presented for the four values of $x$. The curves $a$,
$b$, $c$, $d$ are the polygonal lines connecting the points
with different $k$ and the same $x_{a}$, $x_{b}$, $x_{c}$ and
$x_{d}$, respectively. The line $a$ describes a rotation in vacuum
($x_{a}=0$), and the line $b$ describes a rotation in the continuous
medium ($x_{b}=\infty$) with the dielectric permittivity $\varepsilon=3$
(the Cherenkov condition is satisfied). The calculations were carried out
by the formula (\ref{6}). For simplicity the dependence of $\varepsilon$ on
$k$ (the dispersion) is not taken into account. It followed from the
plots that in a continuous media
        \begin {equation}
        \label{8}
           n_{k}(\infty)\leq\frac{ve^{2}}{hc^{2}}\left(1-\frac{1}
           {\varepsilon\beta^{2}}\right)<\frac{e^{2}}{hc}\approx0.05
        \end {equation}
is larger than the analogous quantity $n_{k}(0)$ in the empty
space. A difference between $n_{k}(\infty)$ and $n_{k}(0)$ is
conditioned by the contribution of the Cherenkov's quanta. Along with
this, the specific oscillations \cite{Tsyt} are revealed on the curve $b$.
They results from the interference of waves in the conditions when the
velocity of the electromagnetic waves propagation is lower than the
velocity of the source motion $c/\sqrt{\varepsilon}<v$.

A similar pattern should be observed also in the case when a medium has
finite sizes. In the section 2 we considered the case of a sphere with
the radius $r_{o}$, around of which electron rotates at the distance
$r_{e}-r_{o}$. The polygonal lines $c$ and $d$ represent the results
of calculations by the formula (\ref{1}) for the two fixed values
$r_{o}/r_{e}=0.974733692=x_{c}$ and $0.980861592=x_{d}$, respectively.
The dielectric permittivity of the sphere $\varepsilon_{0}=3$.
Outside the sphere there is a vacuum ($\varepsilon_{1}=1$). The
electron energy $E_{e}=2MeV$. As it is seen, the specific oscillations
are observed also in this case. However, there are also the peaks, and
on the corresponding harmonics ($k=26$ for the case $c$ and $k=40$
for the case $d$) the radiation is abnormally intensive:
        \begin {eqnarray}
        \label{9}
           n_{26}(x_{c})=4300\qquad\mbox{for the curve }c\,,\nonumber\\
           n_{40}(x_{d})=94\qquad\mbox{for the curve }d.
        \end {eqnarray}
At the same time on the neighbouring harmonics $n_{k}(x)$ is of the
order $n_{k}(\infty)$.

In the empty space the radiation intensity $I_{k}$ reaches a
maximum on the harmonic with $k_{max}=26$: \footnote{This result is
obtained also from the formula
         $k_{max}=0.44(E_{e}/m_{e}c^{2})^{3}$
which is valid for ultrarelativistic electron \cite{Land}.}
         $I_{26}(0)=0.96e^{2}\omega^2_e /c$.
On this harmonic an influence of the sphere with the radius
$r_{o}=0.974733692\ r_{e}$ is the most intensive:
$I_{26}(x_{c})/I_{26}(0)\approx2.53\cdot10^{6}$
(just this value of $r_{o}$ is chosen in the case of the curve $c$). An analogous
situation is possible also on other harmonics. For example, on the
harmonic with $k=40$ an influence of the sphere is maximal at
$r_{o}=0.980861592\ r_{e}$ (the curve $d$). In this case
$I_{40}(x_{d})/I_{26}(0)\approx55700$.

Figs.2 and 3 show the dependence of $n_{k}(x)$ on $x$ for the
harmonics with $k=26$ and $k=40$, respectively. In this
plots also $\varepsilon _{0}=3$, $\varepsilon_{1}=1$ and
$E_{e}=2MeV$. Against a background of the oscillations of the
function $n_{k}(x)$, the extremely narrow and very high peaks are
observed (on the right-hand part the function $n_{k}(x)$ is shown
in the vicinity of the maximal peak). Already at a small deviation
(along the axis of abscissa) from the centre of any of these peaks
$n_{k}$ rapidly decreases. Therefore the value $x=r_{o}/r_{e}$ must
be fixed with a high accuracy (for example, by an external electric
field sustaining a uniform rotation of a particle). The energy radiated
per one period of the electron rotation, is equal to
        \begin {equation}
        \label{10}
           \frac{2\pi}{\omega_{e}}I_{k}=k\hbar\omega_{e}n_{k}.
        \end {equation}

The radiative losses are negligible if the cyclic frequency
        \begin {equation}
        \label{11}
           \omega_{e}\ll \frac{E_{e}}{k\hbar n_{k}}
           \sim10^{13} \frac{E_{e}}{MeV}\frac{10^{8}}{kn_{k}}Hz.
        \end {equation}

An analogous pattern takes place for other $1<\varepsilon_{0}\leq5$ and
$E_{e}\leq5MeV$, when the Cherenkov condition is satisfied (see Table1).
Moreover, in certain cases (see the 2-4th rows of Table 1) one can
observe a superintensive radiation with
        \begin {equation}
        \label{12}
           n_{k}>\frac{2\pi r_{e}}{\lambda_{k}}=k\frac{v}{c}.
        \end {equation}
\\
\bigskip

Table 1: The average number $n_{k}$ of electromagnetic field quanta
emitted per revolution

\hspace {0.6in} of electron. \vspace{0.2in}

\noindent
\begin{tabular} {|l|l|l|l|l|} \hline
\multicolumn{2}{|c|}{\qquad} & Rotation in a continuos medium &
\multicolumn{2}{c|} {Rotation around a sphere in a vacuum} \\
\hline $k$ & $E_{e}$ & \quad $\varepsilon=1$
\qquad $\varepsilon=3$ \qquad\quad  $\varepsilon=5$ &\qquad\qquad $
\varepsilon=3$ \qquad\qquad & $\qquad\quad\varepsilon=5$
\qquad\qquad\\ \cline{3-5}
& MeV & \qquad $n_{k}$ \qquad \qquad $n_{k}$ \qquad \qquad $n_{k}$  &
\qquad $\mu$ \quad\qquad $n_{k}(\mu)$ & $\quad\mu$ \qquad\qquad $n_{k}(\mu)$
\qquad \\
\hline  &\quad$1$ &$3.07\cdot 10^{-4}$ \quad $2.37\cdot 10^{-2}$
\quad $3.18\cdot 10^{-2}$ & $6.6433228$ \qquad $4.13$ & $5.2992$
 \quad \qquad $1.76$ \\
$20$ &\quad$3$ &$2.72\cdot 10^{-3}$ \quad $3.32\cdot 10^{-2}$
\quad $3.33\cdot 10^{-2}$ & $0.5432354$    \qquad $201$ & $3.482$
\ \quad\qquad $0.34$   \\
&\quad$5$ &$3.00\cdot 10^{-3}$ \quad $3.42\cdot 10^{-2}$
\quad $3.63\cdot 10^{-2}$ & $1.480803$ \quad \quad \ $133$ & $2.596109$
\qquad $133$ \\ \hline
&\quad$1$ &$2.39\cdot 10^{-5}$ \quad $1.93\cdot 10^{-2}$
\quad $3.11\cdot 10^{-2}$ &$0.82132$ \quad\quad \ \
\ $9.64$ & $1.13910742$ \quad $2260$   \\
$40$ &\quad$3$ &$1.57\cdot 10^{-3}$ \quad $2.90\cdot 10^{-2}$
\quad $3.77\cdot 10^{-2}$ &$1.2224$ \qquad\quad \ $0.65$ & $0.9986$ \ \
\quad\quad $0.65$   \\
&\quad$5$ &$1.85\cdot 10^{-3}$ \quad $3.22\cdot 10^{-2}$
\quad $3.47\cdot 10^{-2}$ & $4.801$ \qquad\quad \ \ \ $0.16$ & $1.50036$ \
\quad\quad $1.45$ \\ \hline
\end{tabular}\vspace{0.2in}

\noindent
Note: $\varepsilon$ is the dielectric permittivity of the matter. In the
case of a sphere for every three values of $k, E_{e}$ and $\varepsilon$
we chosed and presented one value of the ratio of the sphere radius to
the radius of the electron orbit $r_{o}/r_{e}=1-0.01\mu$, for which
$n_{k}(\mu)$ is considerably larger than $e^{2}/hc$.
\bigskip\\

The formulae (\ref{3}) are not valid for electron rotating inside a spherical
cavity in an infinite medium, and therefore we did not carry out the
corresponding calculations.

The numerical calculations were duplicated by two independent programs.
One of them, a more simple, was made with the help of the Mathematica,
and an another, more fast-acting, on the Pascal language.

\section {Conclusions}

We calculated the intensity of radiation for electron with an energy of
several MeV uniformly rotating around a sphere in its equatorial plane.
The matter of the sphere is regarded as transparent, and its dielectric
permittivity $1<\varepsilon\leq5$. It is obtained that on the average
the $n>k$ quanta of the electromagnetic field may be radiated per
revolution of electron, where $k$ is the number of the radiated harmonic
($k\leq50$). In the absence of a sphere or at the rotation of electron
in an infinite medium with the same $\varepsilon$, the analogous quantity
$n_{k}<0.05\approx e^{2}/hc$. Such an intense radiation takes place when
the sphere surface is at a specific distance from the electron orbit and
when the Cherenkov condition for electron and the matter of the sphere
is satisfied.

The authors are grateful to A. R. Mkrtchyan for the continued
interest to this work and support.

\clearpage

\centerline {\bf Figure captions:}

\bigskip

Fig.1: Average number $n_{k}(x)$ of electromagnetic field quanta emitted
per revolution of electron, as a function of the radiated harmonic's
number $k$. The polygonal lines $a, b, c$ and $d$ differ by the value of
$x$ (the ratio of the sphere radius to the radius of the electron
orbit): $x_{a}=0$ (vacuum), $x_{b}=\infty$ (infinite medium),
$x_{c}\approx 0.9747337$, $x_{d}\approx 0.9808616$. The dielectric
permittivity of the matter $\varepsilon=3$, the electron energy
$E_{e}=2MeV$.

\bigskip

Fig.2: The same quantity, as in Fig.1, depending on $x$. A number of the
radiated harmonic is fixed: $k=26$. Here also $\varepsilon=3$ and
$E_{e}=2MeV$. On the right-hand side the function $n_{k}(x)$ is plotted
in the vicinity of the maximal peak.

\bigskip

Fig.3: The same dependence, as in Fig.2, in the case $k=40$.
\end{document}